\documentclass[12pt]{article}
\usepackage{epsfig}
\usepackage{subfigure}
\usepackage{amssymb,amsmath}
\usepackage{hyperref}
\usepackage{setspace}
\usepackage{url}

\def\tst{\tilde t}
\def\ttau{\tilde \tau}

\hypersetup{
    colorlinks = true,
    linkcolor = blue,
    anchorcolor = blue,
    citecolor = blue,
    filecolor = blue,
    pagecolor = blue,
    urlcolor = blue
}

\begin{document}

\vspace*{1.0cm}

\begin{center}
\baselineskip 20pt {\Large\bf
Coannihilation Scenarios and Particle Spectroscopy in $SU(4)_c \times SU(2)_L \times SU(2)_R$
}

\vspace{1cm}
{\large Ilia Gogoladze\footnote{ E-mail:
ilia@physics.udel.edu\\ \hspace*{0.5cm} On  leave of absence from:
Andronikashvili Institute of Physics, GAS, 380077 Tbilisi, Georgia.}
Rizwan Khalid\footnote{ E-mail: rizwan@udel.edu} and
Qaisar Shafi } \vspace{0.5cm}

{\baselineskip 20pt \it
Bartol Research Institute, Department of Physics and Astronomy, \\
 University of Delaware, Newark, DE 19716, USA \\
}
\vspace{.5cm}

 \vspace{1.5cm} {\bf Abstract}
\end{center}

\baselineskip 18pt

We identify a variety of coannihilation scenarios in a supersymmetric 
$SU(4)_c \times SU(2)_L \times SU(2)_R$
model with discrete left-right symmetry. Non-universal gaugino masses, 
compatible with the gauge symmetry, 
play an essential role in realizing  gluino and bino-wino 
coannihilation regions that are consistent with the
WMAP dark matter constraints. We also explore
regions of the parameter space in which the little hierarchy problem is 
partially resolved. We present several phenomenologically interesting benchmark points
and the associated sparticle and Higgs mass spectra.

\thispagestyle{empty}

\newpage

\addtocounter{page}{-1}


\section{Introduction}

In a recent paper, hereafter referred to as \cite{Gogoladze:2009ug}, we explored the
phenomenology of a supersymmetric $SU(4)_c \times SU(2)_L \times SU(2)_R$ (4-2-2) 
model \cite{pati} with third family Yukawa unification \cite{big-422,bigger-422}. We showed, in 
particular, that Yukawa unification and neutralino dark matter are readily 
compatible in this 4-2-2 model, which should be contrasted with the 
conclusion reached in \cite{Baer:2008jn} within the corresponding 
SO(10) framework where it is extremely hard to do so. The reason for this 
difference largely stems from the 
assumption made in 4-2-2 of tree-level gaugino non-universality, which
is not possible in SO(10). Thus, in contrast to the latter, there is an 
additional soft parameter in 4-2-2 which enables one, via gluino 
coannihilation \cite{Profumo:2004wk}, to make third family Yukawa unification and 
neutralino dark matter mutually consistent. In \cite{Gogoladze:2009ug} we showed some 
benchmark points highlighting the characteristic sparticle and Higgs 
mass spectra in the presence of Yukawa unification. A relatively light gluino is 
a characteristic feature of this 4-2-2 \cite{Gogoladze:2009ug} as well as 
$SO(10)$ \cite{Baer:2008jn} model. 

In this paper we carry out a more thorough investigation 
of the 4-2-2 model without insisting on Yukawa unification. We assume low scale 
gravity mediated supersymmetry breaking \cite{Chamseddine:1982jx}, 
and allow the asymptotic gaugino masses to be non-universal at tree level, compatible 
with the 4-2-2 symmetry \cite{a,b}. We identify a 
variety of coannihilation scenarios in this more general
model which, as in \cite{Gogoladze:2009ug}, also has discrete left-right symmetry
\cite{pati,lr} (more precisely C-parity) \cite{c-parity}. As in 
\cite{Gogoladze:2009ug}, non-universal gaugino masses, compatible with the 4-2-2 gauge 
symmetry, play an essential role in realizing  gluino \cite{Profumo:2004wk} and 
bino-wino \cite{Baer:2005jq} 
coannihilation regions that are consistent with the
WMAP dark matter constraints. We also explore
regions of the parameter space in which the little hierarchy problem is 
partially resolved. We present several benchmark points
and the associated sparticle and Higgs mass spectra. Many of the new particles 
are in a mass range which is accessible at the LHC. 

Supplementing 4-2-2 with a discrete left-right (LR) symmetry reduces the number of 
independent gauge couplings in 4-2-2 from three to two. This is because LR 
symmetry imposes $SU(2)_L$ and $SU(2)_R$ gauge coupling unification condition at 
$M_{\rm GUT}$  ($g_L=g_R$). In 4-2-2 the matter fields are unified into 
three generations of $\psi$ (4, 2, 1), and the antimatter
fields are in three generations of $\psi_c$ $(\bar{4}, 1, 2)$. If the 
minimal supersymmetric standard model (MSSM)
electroweak doublets come from the bi-doublet H(1, 2, 2), the third
family Yukawa coupling H $\psi_c \psi$ yields the following relation
valid at $M_{\rm GUT}$, namely
\begin{equation}
Y_t = Y_b = Y_{\tau} = Y_{\rm Dirac}.
\label{f1}
\end{equation}

In order to get the correct fermion masses and mixings one possibility is to assume 
that the SM Higgs partially comes from $\Delta$ (15, 2, 2) 
dimensional representation of 4-2-2 \cite{Lazarides:1980nt}. 
In other words we have to assume that the SM Higgs doublet is a superposition of $(1,2,2)$ and $(15,2,2)$
Higgs fields. In this case, the superpotential for the Yukawa sector of the 4-2-2 model is
\begin{equation}
Y^{ij}_1\psi_c \psi H + Y^{ij}_2 \psi_c \psi \Delta
\label{f2}
\end{equation}

With the MSSM Higgs doublets now arising from a linear combination of the Higgs 
doublets in (1,2,2) and (15,2,2), the Yukawa unification condition in Eq.(\ref{f1}) 
is in general lost. 

We will assume that due to C-parity the soft mass$^2$ terms, induced
at $M_{\rm GUT}$ through gravity mediated supersymmetry breaking
are equal in magnitude for the scalar
squarks and sleptons of the three families. The tree level asymptotic MSSM
gaugino masses, on the other hand, can be non-universal from the
following consideration. From C-parity, we can expect that the
gaugino masses at $M_{\rm GUT}$ associated with $SU(2)_L$ and $SU(2)_R$
are the same ($M_2 \equiv M_2^R= M_2^L$). However, the 
asymptotic $SU(4)_c$ and consequently $SU(3)_c$ gaugino masses can be 
different. With the hypercharge generator in 4-2-2 given by $Y=\sqrt{\frac{2}{5}}
(B-L)+\sqrt{\frac{3}{5}} I_{3R}$, where $B-L$ and $I_{3R}$ are the
diagonal generators of $SU(4)_c$ and $SU(2)_R$, we have the
following asymptotic relation between the three MSSM gaugino masses:
\begin{equation}
M_1=\frac{3}{5} M_2 + \frac{2}{5} M_3. \label{gaugino-condition}
\end{equation}

The supersymmetric 4-2-2 model with C-parity thus has two
independent parameters ($M_2$ and $M_3$) in the gaugino sector.
The fundamental parameters of the 4-2-2 model are as follows:
\begin{equation}
m_{0}, m_{H_u}, m_{H_d},  M_2, M_3, A_0, \tan\beta, {\rm sign}~{\mu}.
\label{params}
\end{equation}
Thus, compared to the $NUHM2$ model of \cite{Ellis:2008eu}, we have
one additional parameter in 4-2-2 which plays a crucial role in opening up several distinct 
coannihilation channels each associated with a characteristic sparticle and Higgs 
mass spectrum. 

The outline for the rest of the paper is as follows. In Section \ref{constraints_section}
we summarize the scanning procedure and the experimental constraints that we have 
employed. In Section \ref{results} we present the results from our scan, highlight 
some of the predictions of the 4-2-2 model, and display some benchmark points.
Our conclusions are summarized in Section \ref{conclusions}.

\section{Phenomenological constraints and scanning procedure\label{constraints_section}}

We employ the ISAJET~7.78 package~\cite{ISAJET} to perform random scans over
the parameter space listed in Eq.(\ref{params}). In this package, the weak scale values of gauge and
third generation Yukawa couplings are evolved to $M_{\rm GUT}$ via the
MSSM renormalization group equations (RGEs) in the $\overline{DR}$
regularization scheme, where $M_{\rm GUT}$ is defined to be the scale at
which $g_1=g_2$. We do not enforce the unification condition
$g_3=g_1=g_2$ at $M_{\rm GUT}$, since a few percent deviation
from unification can be assigned to unknown GUT-scale threshold
corrections~\cite{Hisano:1992jj}. At $M_{\rm GUT}$, the boundary
conditions are imposed and all the (soft supersymmetry breaking) SSB parameters, along with the gauge
and Yukawa couplings, are evolved back to the weak scale $M_{\rm Z}$.
The impact of the neutrino Dirac Yukawa coupling in the running of the
RGEs is significant only for relatively large values ($\sim 2$ or so) \cite{Gomez:2009yc}.
In our model we expect the largest Dirac coupling to be comparable, at best, 
to the top Yukawa coupling ($\sim 0.5$).

In the evaluation of Yukawa couplings the SUSY threshold corrections~\cite{Pierce:1996zz}
are taken into account at the common scale $M_{\rm SUSY}= \sqrt{m_{\tst_L}m_{\tst_R}}$.
The entire parameter set is iteratively run between $M_{\rm Z}$ and $M_{\rm GUT}$
using the full 2-loop RGEs until a stable solution is obtained. To
better account for leading-log corrections, one-loop step-beta functions
are adopted for gauge and Yukawa couplings, and the SSB parameters $m_i$
are extracted from RGEs at multiple scales $m_i=m_i(m_i)$. The RGE-improved
1-loop effective potential is minimized at an optimized scale $M_{\rm SUSY}$,
which effectively accounts for the leading 2-loop corrections. Full 1-loop
radiative corrections are incorporated for all sparticle masses.

The requirement of radiative electroweak symmetry breaking (REWSB)~\cite{Ibanez:1982fr}
puts an important theoretical constraint on the parameter space. Another
important constraint comes from limits on the cosmological abundance of
stable charged particles~\cite{Yao:2006px}. This excludes regions in the
parameter space where charged SUSY particles, such as $\ttau_1$ or $\tst_1$,
become the lightest supersymmetric particle (LSP). We accept only those
solutions for which one of the neutralinos is the LSP and saturates the WMAP 
dark matter relic abundance bound.

We have performed random scans for the following parameter range:
\begin{eqnarray}
0\leq & m_{0} & \leq 10\, \rm{TeV}, \nonumber \\
0\leq & M_{2} & \leq 1\, \rm{TeV}, \nonumber  \\
0\leq & M_{3} & \leq 1\, \rm{TeV}, \nonumber  \\
-30\, \rm{TeV} \leq & A_{0} & \leq 0, \nonumber \\
0 \leq & m_{Hu} & \leq 10\, \rm{TeV}, \nonumber \\
0 \leq & m_{Hd} & \leq 10\, \rm{TeV}, \nonumber \\
1 \leq & \tan \beta & \leq 65,
\label{ppp1}
\end{eqnarray}

with $\mu >0$ , and $m_t = 172.6$~GeV \cite{Group:2008nq}. A more recent estimate by 
the CDF/D0 collaboration quotes a slightly larger value $m_t = 173.1$~\cite{:2009ec}. This does 
not change our conclusions in any significant way. 

We first collected 1 million points for the 4-2-2 model. To this we added the results of
constrained MSSM (CMSSM) and CMSSM with non-universal Higgs (150,000 points each), as these are subsets of the 4-2-2 model.
All of these points satisfy the requirement of REWSB with the neutralino being the LSP in each case.
Furthermore, all of these points satisfy the constraint $\Omega_{\rm CDM}h^2 \le 10$.
This is done so as to collect more points with a WMAP compatible value of cold dark
matter relic abundance. After collecting the data, we use the
IsaTools package~\cite{Baer:2002fv}
to implement the following phenomenological constraints:
\begin{table}[h]
\centering
\begin{tabular}{rlc}
$m_{\tilde{\chi}^{\pm}_{1}}~{\rm (chargino~mass)}$ & $ \geq\, 103.5~{\rm GeV}$ &  \cite{Yao:2006px}      \\
$m_{\tilde \tau}~{\rm (stau~mass)} $&$ \geq\, 86~{\rm GeV}$                     &   \cite{Yao:2006px}   \\
$m_{\tilde t}~{\rm (stop~mass)} $&$ \geq\, 175~{\rm GeV}$                     &   \cite{Yao:2006px}   \\
$m_{\tilde b}~{\rm (sbottom~mass)} $&$ \geq\, 222~{\rm GeV}$                     &   \cite{Yao:2006px}   \\
$m_{\tilde g}~{\rm (gluino~mass)} $&$ \geq\, 220~{\rm GeV}$                     &    \cite{Yao:2006px}  \\
$m_h~{\rm (lightest~Higgs~mass)} $&$ \geq\, 114.4~{\rm GeV}$                    &  \cite{Schael:2006cr} \\
$BR(B_s \rightarrow \mu^+ \mu^-) $&$ <\, 5.8 \times 10^{-8}$                     &   \cite{:2007kv}      \\
$2.85 \times 10^{-4} \leq BR(b \rightarrow s \gamma) $&$ \leq\, 4.24 \times 10^{-4} \; (2\sigma)$ &   \cite{Barberio:2007cr}  \\
$\Omega_{\rm CDM}h^2 $&$ =\, 0.111^{+0.028}_{-0.037} \;(5\sigma)$               &  \cite{Komatsu:2008hk}    \\
$3.4 \times 10^{-10}\leq (g-2)_{\mu} $&$ \leq\, 55.6 \times 10^{-10}~ \; (3\sigma)$ &  \cite{Bennett:2006fi}
\end{tabular}

\end{table}

We apply the experimental constraints successively on
the data that we acquire from ISAJET. In our plots we exhibit points
that satisfy all of the above constraints as well as those that do not satisfy the 
$(g-2)_{\mu}$ constraint.

\section{Results\label{results}}

\begin{figure}
\centering
\includegraphics{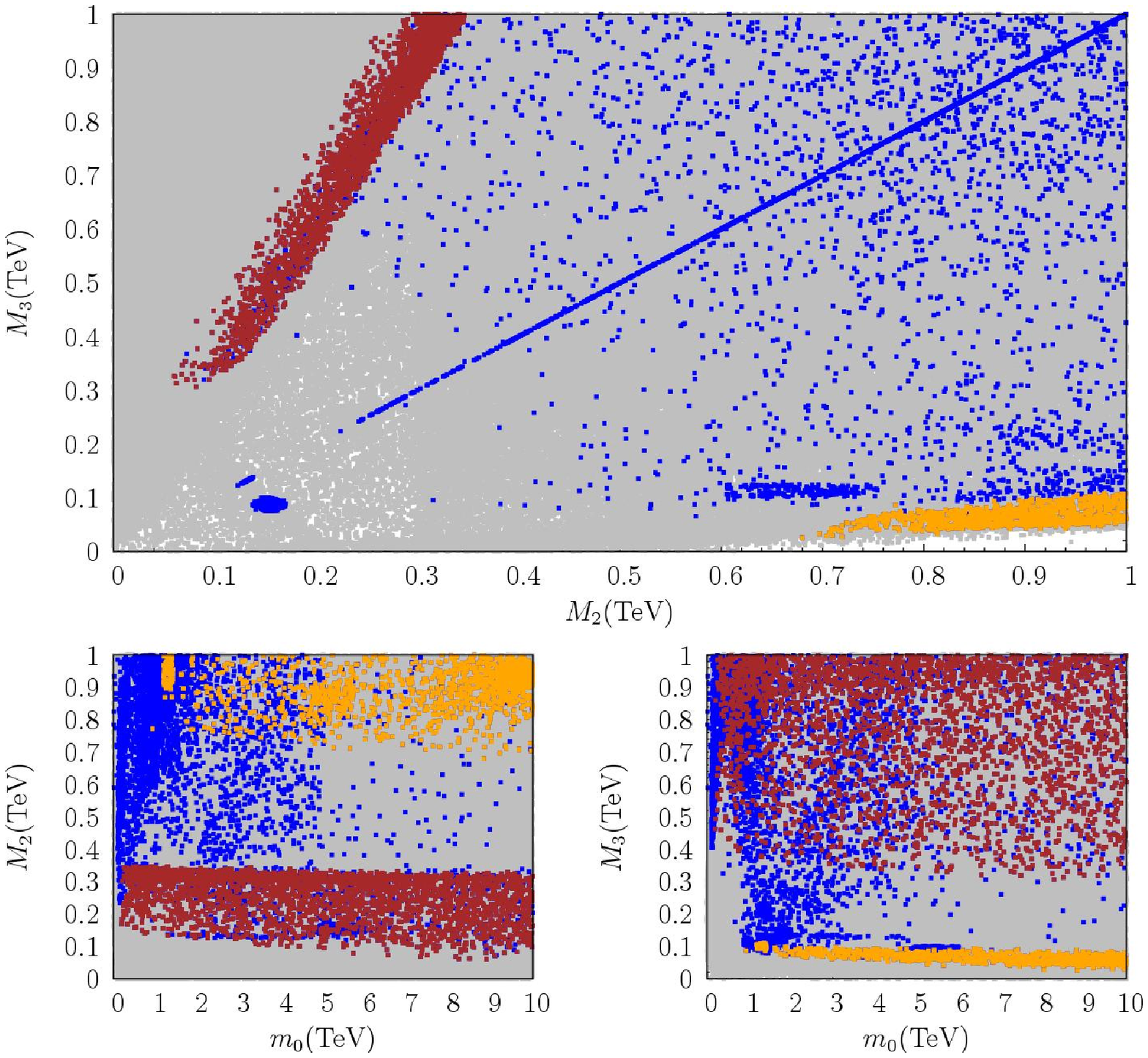}
\caption{
Plots in the $M_2$ - $M_3$, $m_{0}$ - $M_2$ and $m_0$ - $M_3$ planes. Gray points
are consistent with REWSB and $\tilde{\chi}^0_{1}$ LSP. Blue, orange and
red points satisfy the WMAP bounds on $\tilde{\chi}^0_1$ dark matter abundance and
various constraints from colliders ($BR(B_s\rightarrow \mu^+ \mu^-)$,
$BR(b\rightarrow s \gamma)$, and (s)particle mass bounds). Orange points
represent the gluino coannihilation channel, red points depict
the bino-wino coannihilation channel, while blue points represent all other channels.
\label{fund1}}
\end{figure}

\begin{figure}
\centering
\includegraphics{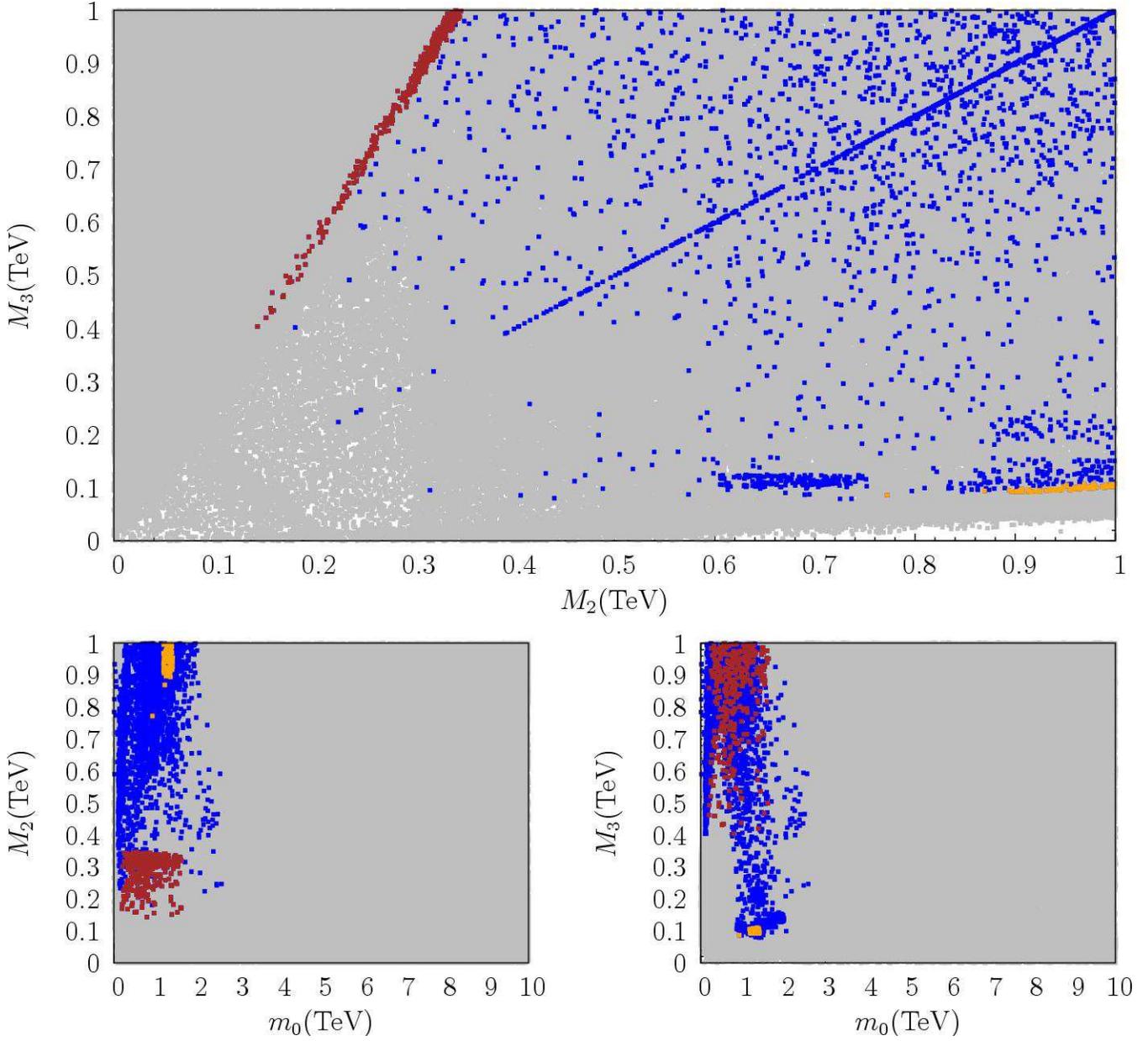}
\caption{
Plots in the $M_2$ - $M_3$, $m_{0}$ - $M_2$ and $m_0$ - $M_3$ planes.
Color coding same as in Figure~\ref{fund1}, except that now
the orange, red and blue points also satisfy the 
$(g-2)_{\mu}$ constraint.
\label{fund2}}
\end{figure}

The asymptotic gaugino mass parameter $M_1$ is constrained by Eq.(\ref{gaugino-condition}) 
at $M_{\rm GUT}$ which is characteristically different from the asymptotic relation 
$M_1=M_2=M_3$ in the case of CMSSM. In the CMSSM, after RGE running, it is therefore not
possible to have the three gauginos nearly degenerate in mass at the low scale.
Indeed at 1-loop level, we have the relations $M_3\sim 6 M_1$ and 
$M_2 \sim 2 M_1 $. In the 4-2-2 model, if we start 
with a low value at $M_{\rm GUT}$ of the ratio $M_3/M_2$, it is 
possible to realize $M_3\sim M_2$ at the TeV scale. 
This scenario offers a relatively light gluino NLSP (next to LSP), so that gluino
coannihilation plays an important role in determining the LSP relic density \cite{Profumo:2004wk}.
Similarly, if we start with a large value of
$M_3/M_2$ at $M_{\rm GUT}$, it is possible to have the wino nearly degenerate with the
bino, in which case one opens up the bino-wino coannihilation channel \cite{Baer:2005jq}. This can be seen
in Figure~\ref{fund1} where we plot results in the $M_2$ - $M_3$, $m_0$ - $M_2$
and $m_0$ - $M_3$ planes. Shown in gray are points that satisfy REWSB and
the requirement that the LSP is a neutralino. The other three colors depict points
that satisfy the above mentioned bounds except the $(g-2)_{\mu}$ constraint.
These include the WMAP bounds on dark matter relic density
and various collider constraints ($BR(B_s\rightarrow \mu^+ \mu^-)$,
$BR(b\rightarrow s \gamma)$, and (s)particle mass bounds). Shown in orange is
the gluino coannihilation channel which occurs at low $M_3$ and high $M_2$ values. We
show the bino-wino coannihilation channel in red which occurs at low $M_2$ and
high $M_3$. Shown in blue are all other coannihilation channels which have been
extensively studied in the CMSSM and CMSSM with non-universal Higgs \cite{Baer:2008ih}. 
Figure~\ref{fund2} is similar to Figure~\ref{fund1} except that we now show the effect of 
imposing in addition the constraint from $(g-2)_{\mu}$. 
The allowed parameter space in this case is appreciably shrunk, which is
understandable because as we increase $m_0$, the MSSM particles gradually decouple from
the theory and we end up with the SM result for $(g-2)_{\mu}$. We note, however,
the two new coannihilation channels introduced by the 4-2-2 model are also present 
in Figure~\ref{fund2}.

\begin{figure}
\centering
\includegraphics{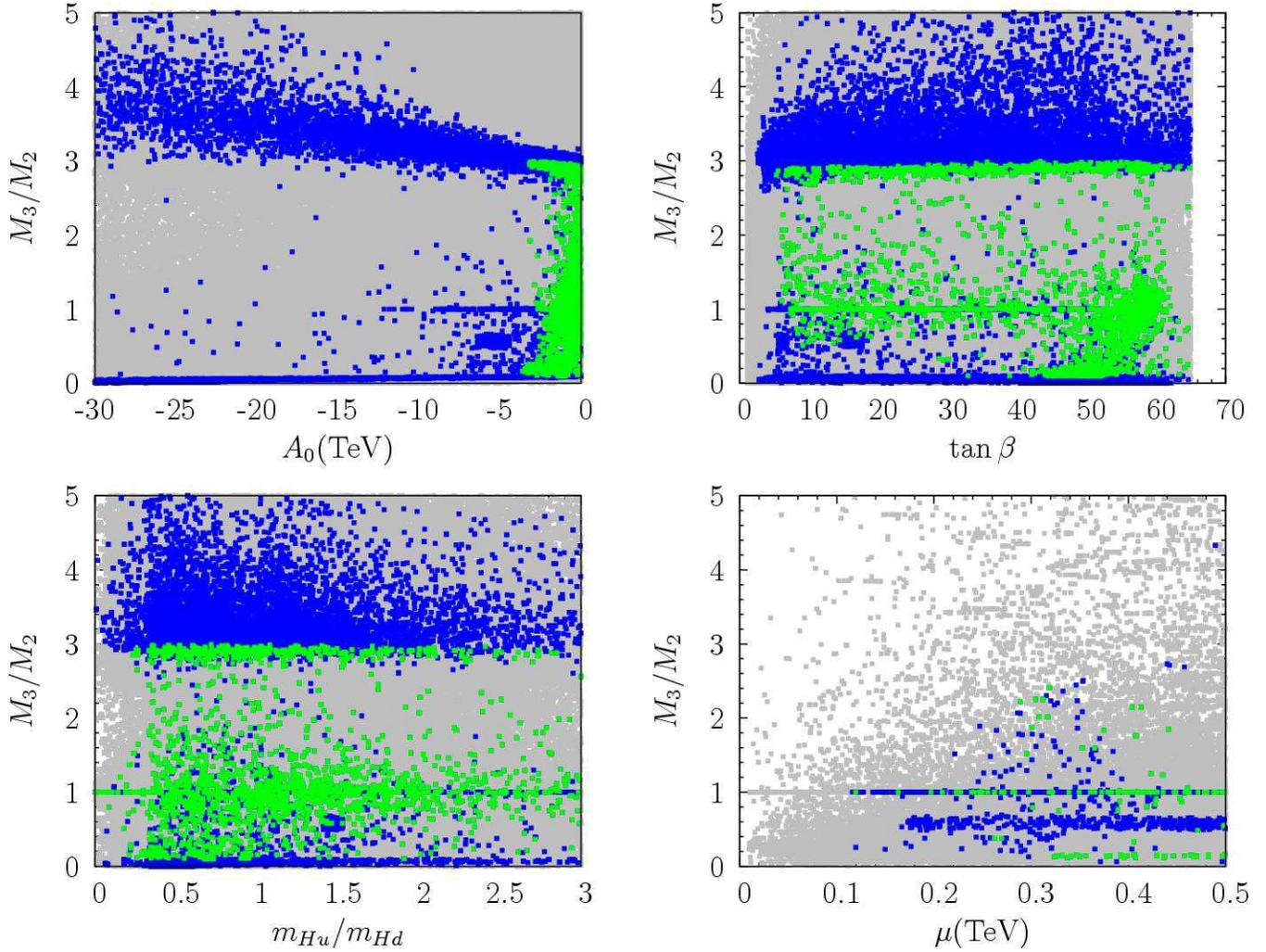}
\caption{
Plots in the $A_0$ - $M_3/M_2$, $\tan\beta$ - $M_3/M_2$,
$m_{Hu}/m_{Hd}$ - $M_3/M_2$ and $\mu$ - $M_3/M_2$ planes.
Gray points are consistent with REWSB and $\tilde{\chi}^0_{1}$ LSP.
Blue points satisfy the WMAP bounds on $\tilde{\chi}^0_1$
abundance and various constraints from colliders
($BR(B_s\rightarrow \mu^+ \mu^-)$, $BR(b\rightarrow s \gamma)$,
and (s)particle mass bounds). Green points also satisfy the
$(g-2)_{\mu}$ constraint.
\label{fund3}}
\end{figure}

In Figure~\ref{fund3} we show results in the $A_0$ - $M_3/M_2$, $\tan\beta$ - $M_3/M_2$,
$m_{Hu}/m_{Hd}$ - $M_3/M_2$ and $\mu$ - $M_3/M_2$ planes. The green points satisfy all the
constraints, while the blue ones satisfy all constraints except the one from $(g-2)_{\mu}$.
We do not distinguish the bino-wino and gluino coannihilation channels in Figure~\ref{fund3}.
The $A_0$ - $M_3/M_2$ plane shows that the $(g-2)_{\mu}$ constraint can only be satisfied
for $|A_0|\lesssim 4 {\rm TeV}$. The $\tan\beta$ - $M_3/M_2$ plane
shows a wide range of allowed values for $\tan \beta$. If we focus
on the $m_{Hu}/m_{Hd}$ - $M_3/M_2$ plane, we can identify points on the $M_3/M_2=1$ line
with the non-universal Higgs model. The CMSSM corresponds to a 
single point in the $m_{Hu}/m_{Hd}$ - $M_3/M_2$
plane, while the $m_{Hu}/m_{Hd}=1$ line displays the effect of varying
the ratio $M_3/M_2$ in the $NUHM1$ parameter space \cite{Ellis:2008eu}, with $M_1$ given by
Eq.(\ref{gaugino-condition}). The results in the $\mu$ - $M_3/M_2$ plane show that
in the 4-2-2 model we have the possibility of realizing relatively small $\mu$ values and, hence,
there exists the potential for ameliorating the little hierarchy problem \cite{Abe:2007kf}. 

\begin{figure}
\centering
\includegraphics{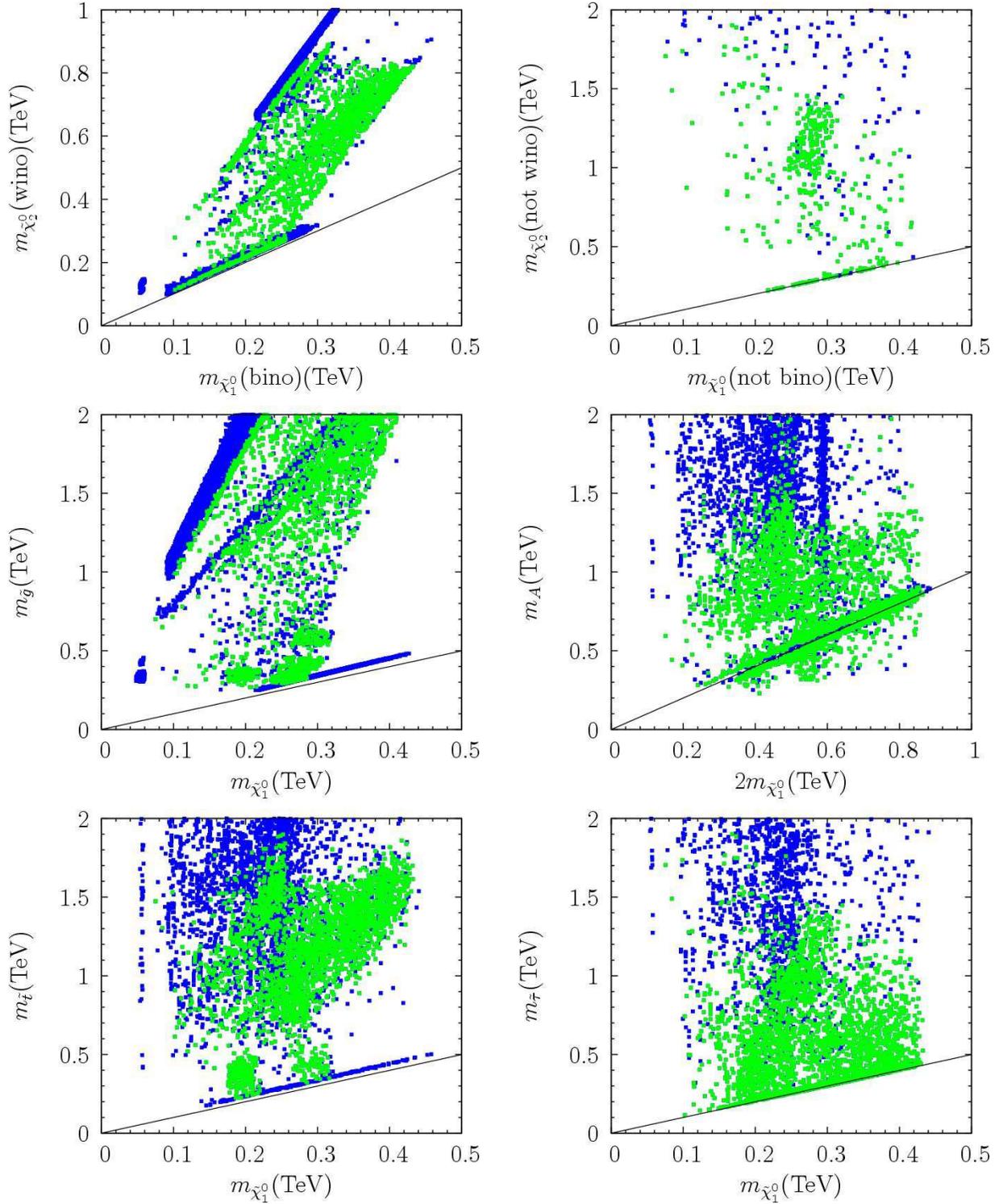}
\caption{
Plots in the $m_{\tilde{\chi}^0_1}(\rm bino)$ - $m_{\tilde{\chi}^0_2}(\rm wino)$,
$m_{\tilde{\chi}^0_1}(\rm not \, bino)$ - $m_{\tilde{\chi}^0_2}(\rm not \, wino)$,
$m_{\tilde{\chi}^0_1}$ - $m_{\tilde g}$, $2 m_{\tilde{\chi}^0_1}$ - $m_{A}$,
$m_{\tilde{\chi}^0_1}$ - $m_{\tilde t}$ and $m_{\tilde{\chi}^0_1}$ - $m_{\tilde {\tau}}$ planes.
Also shown is the unit slope line in each case. Color coding same as in Figure~\ref{fund3}.
\label{spectra1}}
\end{figure}

We exhibit in Figure~\ref{spectra1} the various coannihilation as well as the resonance channels by
plotting the relevant (NLSP) sparticle mass versus the lightest neutralino. The color coding is the same
as in Figure~\ref{fund3} except that we now suppress the gray background corresponding
to the region allowed by REWSB and a neutralino LSP. In the
$m_{\tilde{\chi}^0_1}(\rm bino)$ - $m_{\tilde{\chi}^0_2}(\rm wino)$ plane we show those
solutions for which the lightest neutralino is essentially a pure bino and the NLSP
is primarily a wino. The unit slope line in this plane corresponds to the
bino-wino coannihilation region. In the $m_{\tilde{\chi}^0_1}(\rm not \, bino)$ - 
$m_{\tilde{\chi}^0_2}(\rm not \, wino)$ plane we can recognize the possibility
for coannihilation between the two lightest neutralinos
with the lightest neutralino not a pure bino and the NLSP not primarily a wino. This
case mostly corresponds to the coannihilation of higgsinos. 

We display the gluino coannihilation
region in the $m_{\tilde{\chi}^0_1}$ - $m_{\tilde g}$ plane. Consistent with all known 
bounds, the gluino can be as light as 250 GeV or so. 

We can recognize the $A$-funnel region in the $2 m_{\tilde{\chi}^0_1}$ - $m_{A}$ plane
where neutralino annihilation occurs through the CP-odd Higgs resonance channel.
In the $m_{\tilde{\chi}^0_1}$ - $m_{\tilde t}$ plane
we have the stop coannihilation region with the possibility of a stop as light as
$\sim 200\, {\rm GeV}$. Even though the stop coannihilation region is found in the CMSSM,
the possibility of a light stop consistent with all constraints including the one
from $(g-2)_{\mu}$ is a rather exciting feature of the 4-2-2 model. The stau-coannilation
region is displayed in the $m_{\tilde{\chi}^0_1}$ - $m_{\tilde \tau}$ plane.

\begin{figure}
\centering
\includegraphics{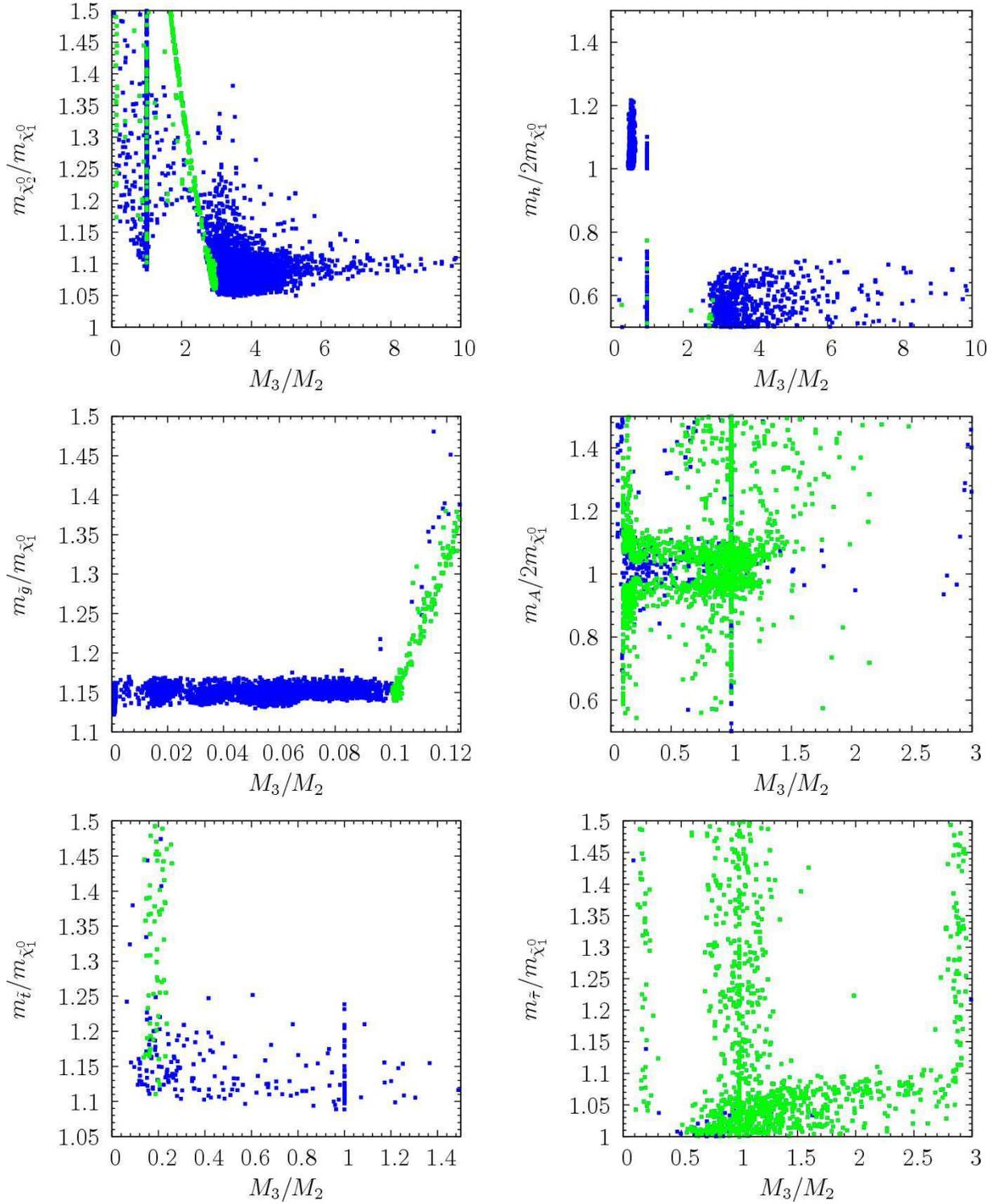}
\caption{
$M_3/M_2$ versus $m_{\tilde{\chi}^0_2}/m_{\tilde{\chi}^0_1}$,
$m_{h}/2 m_{\tilde{\chi}^0_1}$, $m_{\tilde{g}}/m_{\tilde{\chi}^0_1}$, 
$m_{A}/2 m_{\tilde{\chi}^0_1}$, $m_{\tilde{t}}/m_{\tilde{\chi}^0_1}$
and $m_{\tilde{\tau}}/m_{\tilde{\chi}^0_1}$.
Color coding same as in Figure~\ref{fund3}.
\label{spectra1-m3m2}}
\end{figure}

In Figure~\ref{spectra1-m3m2} we show the various coannihilation and resonance channels
as a function of $M_3/M_2$. We have plotted on the y-axis $m_i/m_{\tilde{\chi}^0_1}$ in
the case of coannihilations and $m_i/2 m_{\tilde{\chi}^0_1}$ in the case of resonance
annihilations of the neutralino, where $m_i$ is the NLSP mass. This means that in each figure, we 
need to focus close to the line $y=1$ in order to see coannihilations and resonances. The
$M_3/M_2$ - $m_{\tilde{\chi}^0_2}/m_{\tilde{\chi}^0_1}$ plane shows, once again, that
bino-wino coannihilation occurs for a relatively large value of $M_3/M_2\gtrsim 2.5$. The coannihilations shown
in this plane close to $M_3/M_2 \sim 1$ are due to the higgsinos.
The $M_3/M_2$ - $m_{\tilde{g}}/m_{\tilde{\chi}^0_1}$ plane shows that for gluino
coannihilations to work, one needs $M_3/M_2\lesssim 0.1$ at $M_{\rm GUT}$. It is,
therefore, not surprising that these channels are not found in the CMSSM as they need a
large splitting in the gaugino sector. The other four planes have coannihilation solutions
corresponding to $M_3 \sim M_2$ as well as regions in parameter space that have a splitting 
between $M_2$ and $M_3$. The $M_3/M_2$ - $m_{\tilde{t}}/m_{\tilde{\chi}^0_1}$ plane is rather interesting as it
shows that stop-coannihilation prefers a ratio of $M_3/M_2$ smaller than $1$. This is
understandable because with an increase in the gluino mass, the stop will become heavier. It seems
from the $M_3/M_2$ - $m_{h}/2 m_{\tilde{\chi}^0_1}$ plane that annihilation via the lightest Higgs
resonance occurs for $M_3/M_2\leq1$. However, this may just be due to a lack
of statistics rather than a trend.

Finally we present a few benchmark points in Table~\ref{table1} highlighting phenomenologically 
interesting features of
the 4-2-2 model. Point 1 has been taken from the bino-wino coannihilation region while
point 2 represents a solution in the gluino coannihilation region of the 4-2-2 parameter
space. Gluino coannihilations play a crucial role in reconciling the neutralino dark matter relic density
with Yukawa coupling unification in the 4-2-2 model, as discussed in
\cite{Gogoladze:2009ug}. In Point 3, with a neutralino of mass $\sim 46 \,{\rm GeV}$, 
we demonstrate the possibility of resonant enhancement of neutralino annihilation via the 
$Z$-boson resonance state. Point 4 highlights another 
distinguishing feature of the 4-2-2 model. In the case of the CMSSM,
the left-handed selectron is heavier than the right handed selectron because
of the larger strength of the $SU(2)$ interaction. In 4-2-2, it is possible to achieve
$m_{\tilde {e_L}} < m_{\tilde {e_R}}$. This happens typically for a small $M_3/M_2$ ratio. 
Point 4 corresponds to the bino-wino coannihilation channel, while point 5 represents a 
solution with a relatively small magnitude of $\mu$, so that the little hierarchy 
problem can be ameliorated.

\section{Conclusions\label{conclusions}}

We have shown that the Higgs and sparticle spectroscopy arising from the
supersymmetric $SU(4)_c \times SU(2)_L \times SU(2)_R$ model with non-universal 
gaugino masses can be quite distinct from the well studied constrained 
minimal supersymmetric standard model. Gluino and bino-wino coannihlation 
scenarios in particular, with their associated spectroscopy, are 
particularly exciting possbilities which will be tested at the LHC.

\section*{Acknowledgments}

This work is supported in part by the DOE Grant No. DE-FG02-91ER40626 (I.G., R.K. and Q.S.),
GNSF Grant No. 07\_462\_4-270 (I.G.), and by Bartol Research Institute (R.K.).

\newpage

\begin{table}[t]
\centering
\begin{tabular}{lcccccc}
\hline
\hline
                 & Point 1 & Point 2 & Point 3 & Point 4 & Point 5      \\
\hline
$m_{0}$          & 244.10  & 1193.7  & 4778.5  & 3064.6 & 464.83   \\
$M_{2} $         & 206.02  & 914.41  & 126.07  & 304.23 & 407.81  \\
$M_{3} $         & 563.02  & 93.34   & 66.51   & 927.12 & 492.05  \\
$\tan\beta$      & 13.45   & 49.20   & 7.79    & 52.42  & 12.85  \\
$A_0$            & -26.02  & -1161.5 & -4716.2 & -4928.0& -299.34  \\
$m_{Hu}$         & 156.95  & 381.52  & 4983.6  & 2339.3 & 898.58  \\
$m_{Hd} $        & 134.74  & 534.6   & 4376.1  & 5660.1 & 467.12   \\
${\rm sgn}~\mu$  &  +      &  +      &  +      &   +    &  +   \\
\hline
$m_h$            & 114.7   & 114.7   & 117.9   & 122.8  & 114.9    \\
$m_H$            & 746     & 438     & 4394    & 4783   & 575       \\
$m_A$            & 741     & 435     & 4365    & 4752   & 571      \\
$m_{H^{\pm}}$    & 751     & 447     & 4395    & 4784   & 580      \\
\hline
$m_{\tilde{\chi}^{\pm}_{1,2}}$
                 & 156,742 & 746,946 & 110,402 & 262,2451  & 221,358  \\
$m_{\tilde{\chi}^0_{1,2}}$
                 & 140,155 & 255,745 & 46,109  & 244,262   & 166,230  \\
$m_{\tilde{\chi}^0_{3,4}}$
                 & 736,741 & 928,945 & 390,403 & 2449,2449 & 252,362 \\
$m_{\tilde{g}}$  & 1293    & 295     & 274     & 2207      & 1151  \\
\hline $m_{ \tilde{u}_{L,R}}$
                 & 1157,1159 & 1331,1205 & 4731,4775  & 3540,3392 & 1122,1114   \\
$m_{\tilde{t}_{1,2}}$
                 & 944,1120  & 707,1028  & 2489,3777  & 1855,2385 & 752,1011  \\
\hline $m_{ \tilde{d}_{L,R}}$
                 & 1160,1158 & 1334,1203 & 4732,4748  & 3541,3584 & 1125,1098  \\
$m_{\tilde{b}_{1,2}}$
                 & 1073,1149 & 870,1032  & 3774,4717  & 2360,2562 & 975,1083  \\
\hline
$m_{\tilde{\nu}_{1}}$
                 & 271       & 1334      & 4787       & 2955      & 548  \\
$m_{\tilde{\nu}_{3}}$
                 & 269       & 1181      & 4772       & 1861      & 543  \\
\hline
$m_{ \tilde{e}_{L,R}}$
                & 286,277    & 1336,1216 & 4785,4744  & 2956,3277 & 555,462  \\
$m_{\tilde{\tau}_{1,2}}$
                & 246,308    & 830,1186  & 4714,4769  & 492,1881  & 453,551   \\
\hline
$\mu$           & 730        & 934       & 376        & 2456      & 243    \\
$\Omega_{CDM}h^2$
                & 0.111      &0.105      &0.113       &0.111      & 0.105 \\
\hline
\hline
\end{tabular}
\caption{ Sparticle and Higgs masses (in GeV units),
with $m_t=172.6$ GeV and $\mu>0$. Points 1, 2 and 3 respectively correspond to the bino-wino
coannihilation, gluino coannihilation and $m_Z$ resonance channels. Point 4 
lists the spectrum for a solution for which the left-handed selectron is lighter than the
right-handed selectron. Point 5 is an example of a scenario in which the little-hierarchy problem
is ameliorated. Points 1, 2 and 5 satisfy all the constraints listed at the end of 
Section \ref{constraints_section}, while points 3 and 4 satisfy all constraints except $(g-2)_{\mu}$. 
\label{table1}}
\end{table}

\end{document}